\documentclass[reqno,11pt,a4paper]{amsart}

\usepackage{amssymb}
\usepackage{url}
\usepackage{hyperref}
\usepackage{pdfpages}
\usepackage{graphicx}
\allowdisplaybreaks[4]

\usepackage{mathrsfs}
\let\mathcal\mathscr
\usepackage[mathscr]{eucal}
\usepackage[all]{xy}
\SelectTips{cm}{}

\usepackage{color}

\let\phi=\varphi
\let\kappa=\varkappa

\DeclareMathOperator{\sym}{sym}

\DeclareMathOperator{\const}{const}

\theoremstyle{theorem}
\newtheorem{proposition}{Proposition}

\newtheorem*{Lemma}{Lemma}

\theoremstyle{definition}

\theoremstyle{remark}

\usepackage[all]{xy}
\SelectTips{cm}{}

\usepackage{mathrsfs}
\let\mathcal\mathscr
\usepackage[mathscr]{eucal}

\newcommand{\cprime}{\/{\mathsurround=0pt$'$}}

\begin{document}

\title[Heat transfer equation]{A nonlinear heat transfer equation in turbulent
  media: symmetry classification, recursion operators, and exact solutions}
\author{I.S.~Krasil'shchik}
\address{Trapeznikov Institute of Control Sciences, 65 Profsoyuznaya street,
  Moscow 117997, Russia}
\email{josephkra@gmail.com}
\thanks{Partially supported by the RSF Grant 25-71-20008}
\begin{abstract}
  We study a heat transfer equation in spatial dimensions $n = 1$, $2$,
  and~$3$. A group classification with respect to the functional parameter~$k
  = k(T)$ is done and symmetry algebras are presented. Recursion operators are
  found in the case~$n = 1$ and infinite hierarchies of symmetries are
  constructed. We also find a number of exact solution in all the three cases.  
\end{abstract} 
\date{\today}
\maketitle
\setcounter{tocdepth}{3}
\tableofcontents

\section{Introduction}
\label{sec:introduction}

Heat transfer in turbulent media may be described by the equation
\begin{equation}
  \label{eq:1}
  \frac{\partial T}{\partial t} = 
  k^{-1}\sum_{i = 1}^n
  \frac{\partial^2T}{\partial x_i^2} -
  \frac{n + 2}{2}k^{-2}\sum_{i = 1}^n
  \frac{\partial k}{\partial x_i}\frac{\partial T}{\partial x_i},
\end{equation}
with a functional parameter~$k = k(T)$, $T$ being the temperature,
see~\cite{HT}. The case~$k = k_BT$, where $k_B$ is the Boltzmann constant,
corresponds the delute gases.

We consider the cases $n = 1$, $2$, $3$ and compute symmetry
algebras~$\sym\mathcal{E}$ of Equation~\eqref{eq:1}. These algebras depend on
the form of~$k$ and thus provide a classification of~\eqref{eq:1}.

In the case~$n = 1$, $k = k_0 + k_1T$, $k_i = \const$, the equation admits
recurion operators for symmetries. These operators generate three infinite
hierarchies of symmetries. One of them is local, the other two being
nonlocal. In computation of recursion operators, we use the algorithm
described in~\cite{KVV-Springer}.

In the last section, we present some exact solutions that are invariant with
respect to the symmetries computed in
Sections~\ref{sec:case-n-=}--\ref{sec:case-n-=-1}.

We omit the case~$k = \const$ throughout all the exposition, since it
corresponds to the linear equation.

\section{The case $n = 1$}
\label{sec:case-n-=}

Equation~\eqref{eq:1} takes the form
\begin{equation}\label{eq:2}
 T_t = -\frac{3k_xT_x}{2k^2} + \frac{T_{xx}}{k}
\end{equation}
in this case.

\subsection{Symmetries}
\label{sec:case-n-=-2}

\begin{proposition}\label{prop-1}
  Let the jet order of symmetries be~$\leq 2$. Then\textup{,} depending on the
  form of~$k$\textup{,} the symmetry algebra is generated by the following
  elements\textup{:}
  \begin{description}
  \item[Type 1] for~$k = k_0 + k_1T$\textup{,} $k_i = \const$\textup{,} $k_1\neq 0$,
    \begin{gather*}
      \phi_{10} = T_x,\quad \phi_{11} = \frac{1}{2} k_1 x T_x + k_1 T + k_0,\\
      \phi_{20} = \frac{T_x^2}{(k_1T + k_0)^2}
      - \frac{2T_{xx}}{3 (k_1T + k_0) k_1},\\
      \phi_{21} = \frac{tT_x^2}{(k_1T + k_0)^2}
      - \frac{xT_x}{3k1} - \frac{2tT_{xx}}{3(k_1T + k_0)k_1}.
    \end{gather*}
  \item[Type 2] for $k = (k_0 - T)^{4/5}k_1$
    \begin{gather*}
      \phi_{10} = T_x,\quad \phi_{11} = -\frac{2xT_x}{5} + k_0 - T,\quad
      \phi_{12} = -\frac{x^2T_x}{5} + x(k_0 - T),\\
      \phi_{20} = \frac{T_{xx}}{(k_0 - T)^{4/5}}
      + \frac{6T_x^2}{5(k_0 - T)^{9/5}},\\
      \phi_{21} = \frac{tT_{xx}}{(k_0 - T)^{4/5}}
      + \frac{6tT_x^2}{5(k_0 - T)^{9/5}} + \frac{k_1xT_x}{2}.
    \end{gather*}
  \item[Type 3] $k = (k_0 - T)^{k_2}k_1$, $k_2 = \const$, $k_2\neq 0$, $1$,
    $4/5$,
    \begin{gather*}
      \phi_{10} = T_x,\quad
      \phi_{11} = k_0 - T - \frac{k_2xT_x}{2},\\
      \phi_{20} = \frac{T_{xx}}{(k_0 - T)^{k2}}
      + \frac{3k_2T_x^2}{2(k_0 - T)^{k_2 + 1}},\\
      \phi_{21} = tT_{xx} + \frac{k_1xT_x}{2(k_0 - T)^{k2}}
      + \frac{3k_2tT_x^2}{2(k_0 - T)^{k2 + 1}}.
    \end{gather*}
  \item[Type 4] for $k$ of none of the above forms\textup{,}
    \begin{gather*}
      \phi_{10} = T_x,\quad
      \phi_{21} = \frac{T_{xx}}{k} - \frac{3k_TT_x^2}{2k^2},\quad
      \phi_{22} = \frac{2tT_{xx}}{k} + xT_x
      - \frac{3tk_TT_x^2}{k^2}.
    \end{gather*}
  \end{description}
\end{proposition}

\subsection{Recursion operators}
\label{sec:recursion-operators}

\begin{Lemma}
  Equation~\eqref{eq:2} admits exactly two conservation laws corresponding to
  the cosymmetries
  \begin{equation*}
    g_0 = \frac{1}{\sqrt{k}}, \qquad
    g_1 = \frac{x}{\sqrt{k}}.
  \end{equation*}
  For $k = k_0 + k_1T$ they are~$\omega_i = a_i\,dx + b_i\,dt$\textup{,} $i = 0$\textup{,} $1$\textup{,}
  where
  \begin{gather*}
   a_0 = 2\sqrt{k_1T + k_0}, \quad
    b_0 = \frac{T_xk_1}{(k_1T + k_0)^{3/2}},\\
    a_1 = x\sqrt{k_1T + k_0},\quad
    b_1 = \frac{k_1xT_x}{2(k_1T + k_0)^{3/2}} + \frac{1}{\sqrt{k_1T + k_0}}.
  \end{gather*}
\end{Lemma}
Denote by $\sigma\colon V\to\mathcal{E}$ the two-dimensional covering,
see~\cite{Trends}, that corresponds to the above conservation law and denote
the respective nonlocal variables by~$v^1$ and~$v^2$. Let
also $\mathcal{T}\mathcal{E}$
\begin{align*}
  T_t
  &= \frac{T_{xx}}{k} - \frac{3 k_T T_x^2}{2 k^2},\\
  q_t
  &=  \frac{q_{xx}}{k} - \frac{3k_TT_x q_x}{k^2}
   -\frac{((3kk_{TT} - 6k_T^2) T_x^2 + 2kk_T T_{xx})q}{2k^3}
\end{align*}
be the tangent equation to~\eqref{eq:1}.  Equation~$\mathcal{T}\mathcal{E}$
also admits exactly two conservation laws~$\Omega_i = A_i\,dx + B_i\,dt$, $i =
0$, $1$, with
\begin{gather*}
 A_0 = \frac{q}{\sqrt{k_1 T + k_0}},\quad
B_0 = -\frac{3k_1T_xq}{2 (k_1 T + k_0)^{5/2}} + \frac{q_x}{(k_1 T + k_0)^{3/2}}
,\\
A_1 = \frac{x q}{\sqrt{k_1 T + k_0}},\quad
B_1 = -\frac{(3 k_1x T_x + 2 k_1 T + 2 k_0) q}{2 (k_1 T + k_0)^{5/2}}
+ \frac{x q_x}{(k_1 T + k_0)^{3/2}}.
\end{gather*}
Let $\rho_{\mathcal{T}}\colon W\to \mathcal{T}\mathcal{E}$ be the
corresponding covering with the nonlocal variables~$w_0$ and~$w_1$.

\begin{proposition}\label{prop-2}
  The functions
  \begin{align*}
    \tilde{q}
    &= (xT_x + 2 T) w_0 - T_x w_1
      \intertext{and}
      \tilde{q}
    &= \frac{q_x}{\sqrt{T}} - \frac{3 T_xq}{2 T^{3/2}}
      - \frac{(2 T T_{xx} - 3 T_x^2) w_0}{4 T^2}
  \end{align*}
  together with the defining relations for~$w_0$ and~$w_1$ are recursion
  operator for symmetries of equation at hand.
\end{proposition}
In a more conventional, but less rigorous form these operators can be
presented as
\begin{align}
  \label{eq:3}
  \mathcal{R}_0(\phi)
  &= 2(xT_x +2T)D_x^{-1}(\sqrt{k_1T  + k_0}\cdot\phi)
    - T_xD_x^{-1}(x\sqrt{k_1T  + k_0}\cdot\phi),\\
  \mathcal{R}_1(\phi)
  &= \frac{D_x(\phi)}{\sqrt{T}} - \frac{3T_x\phi}{2T^{3/2}} -
    \frac{2TT_{xx} - 3T_x^2}{2T^2}D_x^{-1}(\sqrt{k_1T  + k_0}\cdot\phi). 
\end{align}
Let us now describe the action of these operators on symmetries
of~$\mathcal{E}$. For simplicity, without loss of generality, we may
assume~$k_0 = 0$, $k_1 = 1$, i.e., $k = T$.

\begin{proposition}
  The operators~$\mathcal{R}_0$ and~$\mathcal{R}_1$ give rise to the following
  hierarchies of symmetries\textup{:}
  \begin{equation*}
    \xymatrix{
      \dots\ar@<.5ex>[r]^-{\mathcal{R}_1}&
      \ar@<.5ex>[r]^-{\mathcal{R}_1}\ar@<1ex>[l]^-{\mathcal{R}_0}\Phi_{-2}^1&
      \ar@<.5ex>[r]^-{\mathcal{R}_1}\ar@<1ex>[l]^-{\mathcal{R}_0}\Phi_{-1}^1&
      \ar@<.5ex>[r]^-{\mathcal{R}_1}\ar@<1ex>[l]^-{\mathcal{R}_0}\Phi_{0}^1&
      \ar@<.5ex>[r]^-{\mathcal{R}_1}\ar@<1ex>[l]^-{\mathcal{R}_0}\phi_{21}&
      \ar@<.5ex>[r]^-{\mathcal{R}_1}\ar@<1ex>[l]^-{\mathcal{R}_0}\phi_{31}&
      \ar@<.5ex>[r]^-{\mathcal{R}_1}\ar@<1ex>[l]^-{\mathcal{R}_0}\phi_{41}&
      \ar@<1ex>[l]^-{\mathcal{R}_0}\dots}
  \end{equation*}
  and
  \begin{equation*}\xymatrixcolsep{7.5mm}
    \xymatrix{
      \dots\ar@<.5ex>[r]^-{\mathcal{R}_1}&\ar@<1ex>[l]^-{\mathcal{R}_0}\Phi_{-1}^2\ar@<.5ex>[r]^-{\mathcal{R}_1}&\Phi_0^2\ar@<.5ex>[r]^-{\mathcal{R}_1}\ar@<1ex>[l]^-{\mathcal{R}_0}&\ar@<1ex>[l]^-{\mathcal{R}_0}\phi_{10}\ar@<.5ex>[dr]^-{\mathcal{R}_1}&&&&&\\
      &&&&\ar@<1ex>[dl]^-{\mathcal{R}_0}\ar@<1ex>[ul]^-{\mathcal{R}_0}0\ar@<.5ex>[r]^-{\mathcal{R}_1}&\phi_{20}\ar@<1ex>[l]^-{\mathcal{R}_0}\ar@<.5ex>[r]^-{\mathcal{R}_1}&\phi_{30}\ar@<1ex>[l]^-{\mathcal{R}_0}\ar@<.5ex>[r]^-{\mathcal{R}_1}&\phi_{40}\ar@<.5ex>[r]^-{\mathcal{R}_1}\ar@<1ex>[l]^-{\mathcal{R}_0}&\ar@<1ex>[l]^-{\mathcal{R}_0}\dots\\
      \dots\ar@<.5ex>[r]^-{\mathcal{R}_1}&\ar@<1ex>[l]^-{\mathcal{R}_0}\Phi_{-1}^3\ar@<.5ex>[r]^-{\mathcal{R}_1}&\ar@<1ex>[l]^-{\mathcal{R}_0}\Phi_0^3\ar@<.5ex>[r]^-{\mathcal{R}_1}&\ar@<1ex>[l]^-{\mathcal{R}_0}\phi_{11}\ar@<.5ex>[ur]^-{\mathcal{R}_1}&&&&&\\
    }
  \end{equation*}
  Thus\textup{,} the operators~$\mathcal{R}_0$ and~$\mathcal{R}_1$ are mutually
  inverse.

  The symmetries
  \begin{gather*}
    \phi_{30} = -\frac{2T_{xxx}}{3 T^{3/2}} + \frac{4T_xT_{xx}}{T^{5/2}}
    - \frac{4 T_x^3}{T^{7/2}},\\
    \phi_{40} = \frac{T_{xxxx}}{6 T^2} - \frac{3 T_x T_{xxx}}{2 T^3}
    - \frac{13 T_{xx}^2}{12 T^3} + \frac{175 T_x^2 T_{xx}}{24 T^4}
    - \frac{21 T_x^4}{4 T^5},\quad \dots
  \end{gather*}
  are local, while all the symmetries~$\phi_{2,i}$, $i\geq2$, and~$\Phi_i^j$,
  $i = 0$, $-1,\dots$, $j = 1$, $2$, $3$, are nonlocal.
\end{proposition}

\section{The case $n = 2$}
\label{sec:sec:case-n-=2}
In the $(1 + 2)$-dimensional case, the symmetries of Equation~\eqref{eq:1} are
described as follows.
\begin{proposition}\label{prop-3}
  Depending on~$k$\textup{,} the symmetry algebra is spanned by the following
  elements
  \begin{description}
  \item[Type 1] $k = k_2\sqrt{T - k_0}$, $k_2\neq 0$,
    \begin{gather*}
      \phi_1 = \frac{T_x^2 + T_y^2}{k_2(T - k_0)^{3/2}} +
      \frac{T_{xx} + T_{yy}}{k_2\sqrt{T - k_0}},\\
      \phi_2 = \frac{t (T_{xx} + T_{yy})}{k_2\sqrt{T - k_0}}
      -\frac{ t (T_x^2 + T_y^2)}{k_2(T - k_0)^{3/2}} - 2 (T - k_0),\\
      \phi_3(f) = f_yT_x + f_xT_y + 4 (T - k_0)f_{xy},
    \end{gather*}
    where $f = f(x, y)$ is an arbitrary function satisfying~$f_{xx} + f_{yy} =
    0$;
  \item[Type 2] $k = k_2(T - k_0)^{k_1}$\textup{,} $k_2\neq 0$\textup{,}
    $k_1\neq 1/2$\textup{,}
    \begin{gather*}
      \phi_1 = T_x,\quad \phi_2 = T_y, \quad \phi_3 = yT_x - xT_y,
      \\
      \phi_4 = x T_x + y T_y + \frac{2(T - k_0)}{k_1},\\
      \phi_5 = \frac{T_{xx} + T_{yy}}{k_2(T - k_0)^{k_1}}
      - \frac{2 k_1 (T - k_0)^{-k1 - 1} (T_x^2 + T_y^2)}{k_2},\\
      \phi_6 = \frac{t (T_{xx} + T_{yy}}{k_2(T - k_0)^{k_1}}
      - \frac{2 t k_1 (T - k_0)^{-k_1 - 1} (T_x^2 + T_y^2)}{k_2}
      - \frac{T - k_0}{k_1}.
    \end{gather*}
  \item[Type 3] general~$k$
    \begin{gather*}
      \phi_1 = T_x, \quad \phi_2 = T_y, \quad \phi_3 = yT_x - xT_y,\\      
      \phi_4 = \frac{T_{xx} + T_{yy}}{k} - \frac{2k_T(T_x^2 + T_y^2)}{k^2},\\
      \phi_5 =\frac{ 2 t (T_{xx} + T_{yy})}{k}
      - \frac{4 tk_T(T_x^2 + T_y^2)}{k^2} + x T_x + y T_y
    \end{gather*}
  \end{description}
\end{proposition}

\section{The case $n = 3$}
\label{sec:case-n-=-1}

Let us finally describe the symmetries of our equation in dimension~$(1 +
3)$. 

\begin{proposition}\label{prop-4}
  Let~$n = 3$. Then the symmetries of~$\mathcal{E}$\textup{,} depending
  on~$k$\textup{,} are spanned by the following elements:
  \begin{description}
  \item[Type 1] $k = k_2(k_0 - T)^{4/11}$, $k_2\neq 0$,
    \begin{gather*}
      \phi_1 = T_x,\quad \phi_2 = T_y,\quad \phi_3 = T_z,\\
      \phi_4 = zT_y - yT_z, \quad \phi_5 = zT_x - xT_z,\quad \phi_6 = yT_x - xT_y,\\
      \phi_7 = xT_x + yT_y + zT_z - \frac{11 (k_0 - T)}{2},\\
      \phi_8 = (-x^2 + y^2 + z^2) T_x - 2 x y T_y - 2 xzT_z + 11 x (k_0 - T),\\
      \phi_9 = (-x^2 - y^2 + z^2) T_z + 2 x z T_x + 2 y z T_y - 11 z (k_0 - T),\\
      \phi_{10} = (-x^2 + y^2 - z^2) T_y + 2 x y T_x + 2yz T_z - 11 y (k_0 - T),\\
      \phi_{11} = \frac{10 (T_x^2 + T_y^2 + T_z^2)}{11k_2(k_0 - T)^{15/11}}
      + \frac{T_{xx} + T_{yy} + T_{zz}}{k_2(k_0 - T)^{4/11}},\\
      \phi_{12} = \frac{t (T_{xx} + T_{yy} + T_{zz})}{k_2(k_0 - T)^{4/11}}
      +\frac{10 t (T_x^2 + T_y^2 + T_z^2)}{11k_2(k_0 - T)^{15/11}}
      + \frac{11 (k_0 - T)}{4}.
    \end{gather*}
  \item[Type 2] $k = k_2(k_0 - T)^{k_1}$\textup{,}
    $k_1\neq 0\textup{,} 4/11$\textup{,} $k_2\neq 0$\textup{,}
    \begin{gather*}
      \phi_1 = T_x,\quad \phi_2 = T_y,\quad \phi_3 = T_z,\\
      \phi_4 = zT_y - yT_z,\quad \phi_5 = zT_x - xT_z,\quad \phi_6 = yT_x - xT_y,\\
      \phi_7 = x T_x + y T_y + zT_z -\frac{ 2 (k_0 - T)}{k_1},\\
      \phi_8 = \frac{5k_1 (k_0 - T)^{-k_1 - 1}(T_x^2 + T_y^2 + T_z^2)}{2k_2}
      + \frac{T_{xx} + T_{yy} + T_{zz}}{k_2(k_0 - T)^{k_1}},\\
      \phi_9 = \frac{t (T_{xx} + T_{yy} + T_{zz})}{k_2(k_0 - T)^{k_1}}
      +\frac{5 t k_1 (k_0 - T)^{-k_1 - 1} (T_x^2 + T_y^2 + T_z^2)}{2k_2}
 + \frac{k_0 - T}{k_1}.
    \end{gather*}
  \item[Type 3] $k$ of the general form,
    \begin{gather*}
      \phi_1= T_x,\quad \phi_2= T_y,\quad \phi_3= T_z,\\
      \phi_4= zT_y - yT_z,\quad \phi_5= zT_x - xT_z,\quad \phi_6= yT_x - xT_y,\\
      \phi_7 = \frac{T_{xx} + T_{yy} + T_{zz}}{k}
      - \frac{5k_T(T_x^2 + T_y^2 + T_z^2)}{2k^2},\\
      \phi_8 = \frac{2t(T_{xx} + T_{yy} + T_{zz})}{k}
      - \frac{5tk_T(T_x^2 + T_y^2 + T_z^2)}{k^2}
      + x T_x + y T_y + zT_z.
    \end{gather*}
  \end{description}
\end{proposition}

\section{Exact solutions}
\label{sec:exact-solutions}

Unfortunately, the capacities of our software are rather limited, so we
managed to find quite a few exact symmetry-invariant solutions of
Equation~\eqref{eq:1}. They are presented in the forthcoming subsections.

Everywhere below, $\alpha$'s, $\beta$'s, and~$\gamma$'s are constants.

\subsection{$n = 1$, $k = T$}
\label{sec:n-=-1}

The $\phi_{11}$-invariant solution is
\begin{equation*}
  T(x, t) = \frac{\alpha}{x^2}.
\end{equation*}
The $\phi_{30}$-invariant solution is
\begin{equation*}
  T(x, t) = \frac{2}{\alpha_1((x + \alpha_3)^2 - \alpha_2\exp(\alpha_1t))}.
\end{equation*}
There are two travelling-wave solutions presented in the implicit form:
\begin{gather*}
  \frac{ 4 \mu^2\ln{(2\sqrt{\alpha_1 T(\tau)} - \alpha_1)}}{\alpha_1}
  - \frac{2 \mu^2 \ln(\alpha_1 T(\tau))}{\alpha_1}
  + \frac{2\mu^2}{\sqrt{\alpha_1T(\tau)}} - \tau - \alpha_2 = 0
  \intertext{and}
  \frac{4 \mu^2\ln(2\sqrt{\alpha_1T(\tau)} + \alpha_1)}{\alpha_1}
  - \frac{2 \mu^2}{\sqrt{\alpha_1T(\tau)}}
  - \frac{2 \mu^2 \ln(\alpha_1T(\tau))}{\alpha_1} - \tau - \alpha_2 = 0,
\end{gather*}
where $\tau = \mu x + t$.

\subsection{$n = 2$, $k = \sqrt{T}$}
\label{sec:n-=-2}

The general travelling-wave solution is given by the integral
\begin{equation*}
  \int^{T(\tau)}
  \frac{d\lambda}{\lambda C_1+\dfrac{2\lambda^{3/2}}{\alpha^2+\beta^2}}
  - \tau - C_2 = 0,
\end{equation*}
where $\tau = t + \alpha x + \beta y$. For $C_1 = 0$ we have a simple solution
\begin{equation*}
  T(\tau) = \frac{(\alpha^2 + \beta^2)^2}{(C_2 + \tau)^2}.
\end{equation*}
If $C_1\neq 0$, the solution is presented in the implicit form
\begin{multline*}
  -\ln\big(-(\alpha^2 + \beta^2)^2 C_1^2 + 4 T(\tau)\big)
  + \ln\big(-\alpha^2 C_1 - \beta^2 C_1 + 2 \sqrt{T(\tau)}\big)\\
  - \ln\big(2\sqrt{T(\tau)} + C_1 (\alpha^2 + \beta^2)\big)
  + \ln(T(\tau)) + C_2 - \tau C_1 = 0
\end{multline*}

Here are some $t$-independent rotation-invariant solutions:
\begin{equation*}
  F(r) = C_2\exp\Big(\int^{T(r)}
  \frac{\exp(-\lambda)\,d\lambda}{(\exp(-\lambda)\lambda - C_1)}\Big)
\end{equation*}
and
\begin{equation*}
  T(r) = \frac{C_1}{r^4},\qquad T(r) = C_2r^2,
\end{equation*}
where $r = \sqrt{x^2 + y^2}$.

\subsection{$n = 3$, $k = T^{4/11}$}
\label{sec:n-=-3}

The general travelling-wave solution is given by the integral formula
\begin{equation*}
\int^{T(\tau)}\dfrac{d \lambda}{C_1\lambda^{\dfrac{10}{11 \alpha^2 + \beta^2 +
      \gamma^2}}
  + \dfrac{11 \lambda^{15/11}}{5 (3 \alpha^2 + 3 \beta^2 + 3 \gamma^2 - 2)}} - \tau - C_2 = 0,
\end{equation*}
where $\tau = t + \alpha x + \beta y + \gamma z$. In particular, when $C_1 =
0$, one has
\begin{equation*}
  T(\tau) = \left(
    \frac{130 (C2 + \tau) (3\alpha^2 + 3\beta^2 + 3\gamma^2 - 2)}{121}
  \right)^{11/26}.
\end{equation*}
We found also two rotation-invariant solutions:
\begin{equation*}
  T(r) = \left(\frac{C_1r - 2C_2}{11r}\right)^{11}\text{ and }
  T(r) =  -\left(\frac{C_0 + C_1t}{r^2}\right)^{11/4},
\end{equation*}
where $r = \sqrt{x^2 + y^2 + z^2}$.

\section*{Acknowledgements}
\label{sec:acknowledgements}

It is my pleasure to thank Valentin Lychagin for discussions.

Symbolic computations were done by the \textsc{Jets} software,~\cite{BM-Jets}.


\begin{thebibliography}{19}
\bibitem{BM-Jets} H.~Baran, M.~Marvan, \emph{Jets. A software for differential
    calculus on jet spaces and diffieties}.  Software Guide, Silesian
  University, Opava, Czech Republic, Jets 4, \url{http://jets.math.slu.cz/}

\bibitem{KVV-Springer} Joseph Krasil{\cprime}shchik, Alexander Verbovetsky,
  Raffaele Vitolo, \emph{The Symbolic Computation of Integrability Structures
    for Partial Differential Equations}. Texts \& Monographs in Symbolic
  Computation, Springer, 2017.
  
\bibitem{HT} Valentin Lychagin,
\emph{On geometry of turbulent flows},
Journal of Geometry and Physics,
Volume 217,
2025,
105646,
\url{https://doi.org/10.1016/j.geomphys.2025.105646}.
  
\bibitem{Trends} I.S.\ Krasil{\cprime}shchik, A.M.\ Vinogradov. Nonlocal
  trends in the geometry of differential equations: Symmetries, conservation
  laws, and B\"{a}cklund transformations. Acta Appl Math \textbf{15}, 161--209
  (1989). \url{https://doi.org/10.1007/BF00131935}
  
\end{thebibliography}
\end{document}